# High energy ball-milling and Synthesis temperature study to improve superconducting properties of MgB2 ex-situ tapes and wires.

| Journal: | *Applied Superconductivity Conference* |
|---|---|
| Manuscript ID: | draft |
| Subject Area: | Materials |
| Date Submitted by the Author: | n/a |
| Complete List of Authors: | Romano, Gennaro; CNR-INFM-LAMIA and Università di Genova<br>vignolo, maurizio; CNR-INFM-LAMIA<br>braccini, valeria; CNR-INFM-LAMIA<br>malagoli, andrea; CNR-INFM-LAMIA<br>bernini, cristina; CNR-INFM-LAMIA<br>tropeano, matteo; CNR-INFM-LAMIA<br>fanciulli, carlo; CNR-INFM-LAMIA<br>putti, marina; CNR-INFM-LAMIA |

# High energy ball-milling and Synthesis temperature study to improve superconducting properties of MgB$_2$ *ex-situ* tapes and wires

Gennaro Romano, Maurizio Vignolo, Valeria Braccini, Andrea Malagoli, Cristina Bernini, Matteo Tropeano, Carlo Fanciulli, Marina Putti and Carlo Ferdeghini

*Abstract*— MgB$_2$ monofilamentary nickel-sheated tapes and wires were fabricated by means of the *ex-situ* powder-in-tube method using either high-energy ball milled and low temperature synthesized powders. All samples were sintered at 920°C in Ar flow. The milling time and the revolution speed were tuned in order to maximize the critical current density ($J_c$) in field: the maximum $J_c$ value of 6 x 10$^4$ A/cm$^2$ at 5K and 4T was obtained corresponding to the tape prepared with powders milled for 144h at 180rpm. Various synthesis temperature were also investigated (730-900°C) finding a best $J_c$ value for the wire prepared with powders synthesized at 745°C. We speculate that this optimal temperature is due to the *fluidifying* effect of unreacted magnesium content before the sintering process which could better connect the grains.

## I. INTRODUCTION

A great deal of work on MgB$_2$ development has been done by many groups since MgB$_2$ was discovered [1], and significant progress is being made in improving basic properties such as the critical current ($J_c$), the upper critical field ($H_{c2}$) and the irreversibility field ($H_{irr}$). Most of these efforts have been focused on the improvement of the high-field J$_c$ and $H_{c2}$ by chemical doping with carbon (C) containing compounds, such SiC, C, B$_4$C and carbon nanotube (CNTs) [2]-[7].

However, the doping effects have been limited by the agglomeration of nanosized dopants and poor reactivity between boron and C.

Moreover, the self-field and the low-field $J_c$ were depressed due to the decrease in the superconducting volume.

To improve these properties, various methods have been reported. They include, for example, a ball-milling method [8], [9], excess Mg addition [10], low-temperature solid-state reaction [11] and the use of MgH$_2$ powder instead of Mg [12].

The majority of these works have been made on bulks and on *in-situ* wires and tapes so, no systematic study of the effects of ball-milling parameters and synthesis temperature on *ex-situ* conductors has been performed yet.

The *ex-situ* technique permits the development of long conductors and multifilamentary tapes easier than *in-situ* method [13], allows a better control of the granulometry and purity degree of the starting powders, and the conductors obtained are more homogeneous.

However, the $J_c$ behavior in magnetic field for the samples prepared by the *ex-situ* way is not so good as for the *in-situ* samples. Therefore, a further development of the starting MgB$_2$ powders is needed to make the *ex-situ* conductors definitely competitive.

In this paper, we present the results obtained by changing the properties of the starting MgB$_2$ powders on the $J_c$ vs B behavior by two ways: varying high energy ball-milling parameters and varying the synthesis temperatures.

## II. EXPERIMENTAL DETAILS

MgB$_2$ powders were prepared from commercial amorphous B (95-97% purity) and Mg (99.99% purity): the powders were mixed and underwent a heat treatment at various temperature between 730 and 900°C. Powders synthesized at 900°C were high energy ball-milled in a planetary ball mill with WC jar and media for different milling times (0-256 h) and with a milling revolution speed of 180 rpm. In order to reduce the oxygen contamination the whole process was performed in glove box with a directly linked oven and using high purity Ar flow: a set of samples with powders synthesized between 730 and 900°C was realized in controlled atmosphere. The powders synthesized at 900°C were also milled with revolution speed between between 240 and 360 rpm for 36 and 72 h.

Monofilamentary tapes and wires were fabricated by means of the *ex-situ* powder-in-tube technique [14]. MgB$_2$ powders were packed inside Ni tubes. The tubes were grove rolled and drawn down to a diameter of 1.4 mm. We used this configuration for the synthesis temperature study. For the high energy ball-milling study the wire was then cold rolled in several steps to a tape of about 0.35 mm in thickness and 4 mm in width.

Manuscript received 19 August 2008. This work was supported in part by the Columbus Superconductors S.p.A and by MIUR under the project FIRB-MAST (RBIP06M4NJ). We thank also the Compagnia di S. Paolo for the financial support.

G. Romano, M Tropeano, C. Fanciulli, M. Vignolo, A. Malagoli, V. Braccini, C. Bernini, M. Putti and C. Ferdeghini are with CNR-INFM LAMIA, C.so Perrone 24, 16152 Genova, Italy (phone: +39-010-6598789; fax: +39-010-6598732; e-mails: maurizio.vignolo@infm.it, romano@fisica.unige.it, malagoli@lamia.infm.it, carlo.fanciulli@infm.it, braccini@lamia.infm.it; cristina.bernini@infm.it; tropeano@fisica.unige.it, amartin@chimica.unige.it; putti@fisica.unige.it, ferdeghini@fisica.unige.it

The superconducting transverse cross section of the conductor was 0.6 mm$^2$ for the wire and 0.2 mm$^2$ for the tape configuration. All samples were sintered at 920°C in Ar flow.

The microstructure of MgB$_2$ powders was investigated using a Scanning Electron Microscope (SEM) with which the grain size distribution was carried out, and X-ray diffraction (XRD) to identify the phase and to estimate the crystallite size.

Short pieces of about 6 mm in length cut from the Ni-sheathed tapes and wires were employed to perform magnetization measurements vs. magnetic field in a commercial 5.5 T MPMS Quantum Design Squid magnetometer. The magnetic field was applied perpendicular to the tape surface and critical current density values were extracted from the *M-H* loops applying the appropriate critical state model. Demagnetization corrections are negligible at field above 0.7 T, being the Nickel saturation field [15].

### III. RESULTS AND DISCUSSIONS

It is well known that the critical current density of MgB$_2$ *ex-situ* conductors is directly influenced by the micro structural properties of the initial superconducting powders.

The aim of this work is to study how optimize these properties to improve the critical current density behaviour of the superconducting tapes and wires following two ways: by varying the milling time and/or energy in a high energy ball-milling process or by varying the powders synthesis temperature.

*Critical current improvement through high energy ball-milling*

The ball-milling effect is well known: it is used to reduce the powders average grain size to improve the performances of the samples in magnetic field.

In our previous work [16]-[18] we have shown how $J_c$ increases when the milling time and/or the milling revolution speed increases (for a detailed study on ball-milling parameterization see also the work by Mio [19]). In both cases we obtained an improvement until a maximum value is reached and then we observed a decrease for higher milling times and/or speeds.

Here we perform a more detailed study and show how this feature is correlated to the empirical parameter which we will call *"Milling Parameter"* (MP); i.e. the product between the square of the milling revolution speed (rpm$^2$), proportional to the milling energy, and the milling time (h).

Our data are summarized in Fig.1a where the critical current density as a function of the magnetic field is shown at 5 K as extracted from the *M-H* loops with the field perpendicular to the tape surface for the entire set of samples. For simplicity the symbols used in Fig.1a are the same for all the subsequent plots.

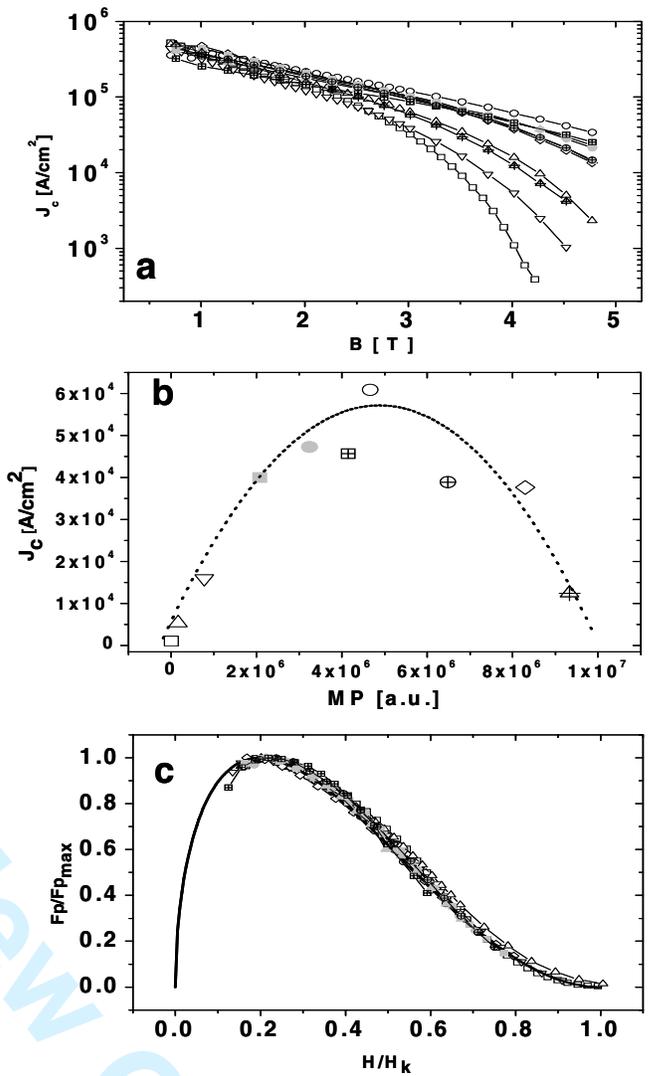

Fig. 1 (a) Magnetic Jc vs B; (b) Magnetic Jc vs MP at 4T,5K; (c) Normalized pinning force vs normalized field. Empty symbols: milled 0-144-24-5-256 h x 180rpm; Full symbols: milled 240-300-360 rpm x 36 h; Cross symbols: milled 240-300-360 rpm x 72 h.

To better visualize this behaviour we plot in Fig.1b the $J_c$ value obtained at 4T and 5K as a function of MP (the line is only a guide for the eye). We obtained the maximum $J_c$ value of 6 x 10$^4$ A/cm$^2$ corresponding to the optimal milled tape prepared with powders synthesized at 900°C and milled for 144h at 180rpm.

Quite similar results can be obtained for the tape whose powders are milled for 36h at 300rpm, thus strongly reducing the milling time.

In our previous work [17] we have shown, by means of magnetization measurements in magnetic field, parallel and perpendicular to the tape surface, that the critical current anisotropy already disappear for this degree of milling.

In Fig.1c the normalized pinning force at 5K is reported as a function of the normalized irreversibility field $H/H_{irr}$. The irreversibility field was determined with a Kramer linear extrapolation and is indicated in the plot as the Kramer field $H_{irr}=H_k$. For comparison also the theoretical curve of grain boundary pinning model, with its maximum at 0.2, is reported.

As we can clearly see, the pinning force behavior is well described by the grain boundary based pinning model and is the same for all the samples, suggesting that the $J_c$ changes cannot be explained in terms of modifications of the pinning mechanism.

In Fig.2a the average particle size obtained from SEM analysis is plotted as a function of MP. The effect of ball milling is evident already for low MP values, where the particle size abruptly decreases from 1.4 to 0.4 µm. Further increasing MP has only a weak effect on the particle size, which remain almost constant between 0.3 and 0.4 µm.

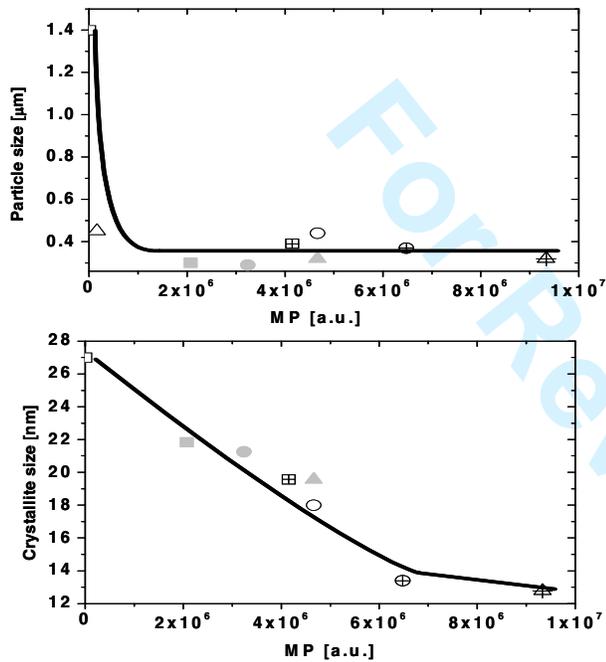

Fig. 2. (a) Average particle size obtained from SEM analysis; (b) Average crystallite size obtained from XRD analysis. The lines are only a guide for the eye.

An interesting result was obtained going to investigate the average crystallite size. In Fig.2b is shown the lower limit of the average crystallite size calculated by the Debye-Scherrer formula, as a function of MP. The values are determined from the half width of the (002) diffraction peak using the formula $D = \alpha \lambda / \beta \cos\theta$ where $D$ is the mean crystallite size, $\alpha$ is a geometrical factor (= 0.94), $\lambda$ is the X-Ray wavelength (=1.54056 Å), $\beta$ is the half width of the diffraction peak, and $\theta$ the angular position of the diffraction peak.

These values are lower than those obtained for the particle size from SEM analysis, ranging from 13 to 27 nm, suggesting that the grains are polycrystalline agglomerate of smaller crystallite and that ball milling continues to be effective increasing MP.

A complete characterization of the optimal milled tape has been done in our recent work [18] where $H_{c2}$, resistivity and magneto optical measurements have been performed. For this tape, the perpendicular critical field at 25K increases from 6T (for the not milled sample) to 9.5T. Our analysis correlate the $J_c$ in-field improvement to the enhanced $H_{c2}$ value due to extra electrons scattering coming from the reduced crystallite size, comparable to the mean free path of $MgB_2$.

Nevertheless, at higher MP the disorder increases, as in C-doped or in neutron irradiated sample [20] and this may be the reason why $J_c$ decreases for higher MP values.

*Critical current improvement through synthesis temperature*

As we said in the introduction, most of the studies on the synthesis of $MgB_2$ are done for bulks or for *in-situ* tapes or wires.

Here we report on the behaviour of the critical current density of $MgB_2$ *ex-situ* wires as a function of synthesis temperature. We explored various temperatures ranging from 730 to 900°C synthesizing and managing the powders in a glove box, to avoid any oxygen contamination effects, until the powders were sealed into the tube. The oxygen content is in fact a crucial parameter for what concern the $J_c$ in-field behavior of the samples and can affect the pinning mechanism [21].

The main and substantial difference respect to bulk and *in-situ* conductors fabrication process is that the powders undergo two heat treatments.

The first heat treatment is analogous to the bulk synthesis process while the second happen at the end of the mechanical deformation process in order to sinter the grains between them and relax the powders and the sheath strain [14], [22].

In Fig.3 is shown the magnetic $J_c$ at 5K as a function of the perpendicular applied magnetic field for the wires synthesized at various temperatures. For simplicity the symbols used in Fig.3 are the same for all the subsequent plots.

We remind here that these samples are square wires and they therefore show a better in-field $J_c$ respect to the flat tapes of Fig.1a.

For some samples the data at lower field are not reported because of the presence of flux-jumps in the hysteresis loop due to magnetic and/or thermal instabilities. Even without these data points it is evident that the sample synthesized at high temperature (900°C) has a lower $J_c$ than of those synthesized at low temperatures (around 745°C).

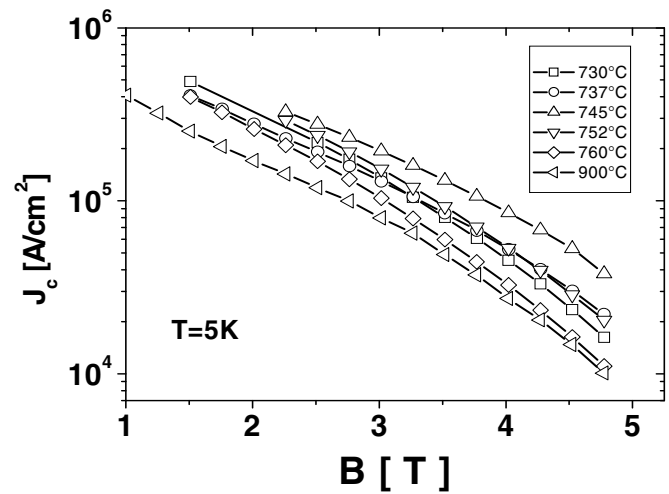

Fig. 3. Magnetic Jc vs B for the wires synthesized at various temperature

The sample with the highest $J_c$ is obtained for the synthesis temperature value of 745°C.

To better visualize the effect of synthesis temperature we plot in Fig.4 the $J_c$ value obtained at 4T and 5K as a function of synthesis temperature.

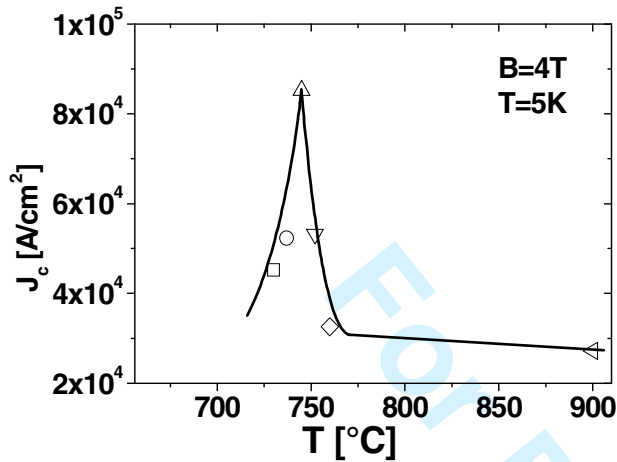

Fig. 4. Magnetic Jc vs synthesis temperature at 4T, 5K. The line is only a guide for the eye

The critical current density first increases slowly, decreasing the temperature from 900 to 752°C and then rapidly increases reaching a maximum at 745°C. Further reducing T gets $J_c$ worse again.

In Fig.5 the magnesium relative content is shown as a function of the synthesis temperature. This can be roughly estimated from the ratio between the XRD intensities of the (101) peak of $MgB_2$ and the (220) peak of Mg for all the samples.

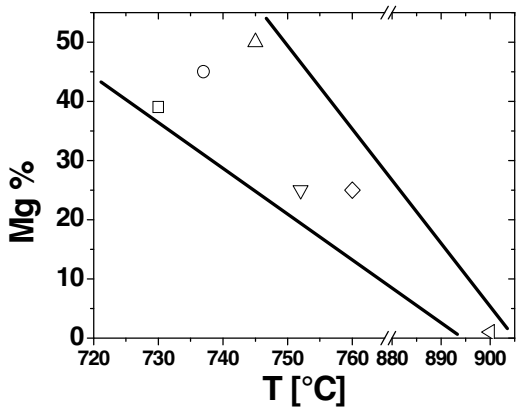

Fig. 5. Relative magnesium content vs synthesis temperature. The lines are only a guide for the eye.

As foreseen, the amount of unreacted Mg increases by lowering the synthesis temperature therefore decreasing the amount of reacted $MgB_2$.

We speculate that the wire synthesized at high temperature (900°C) has a lower $J_c$ with respect to the others because the absence of Mg and B during the mechanical deformation process is detrimental to a correct grain packaging.

The presence of the latter could act as a *fluidifying* for the $MgB_2$ grains, resulting therefore more connected. This could happen because during the deformation process, where the grains are subjected to stress and cracks, the presence of Mg and B can fill the voids left from the imperfect packaging of broken grains.

The subsequent sintering at 920°C then, has the double aim to complete the chemical reaction between the unreacted Mg and B and to finally sinter and relax the whole grains.

It is a middle way between the *in-situ*, where the deformation can pre-activate some Mg and B atoms to form $MgB_2$ before the final sintering, and a *full ex-situ* process, where the $MgB_2$ phase is completely formed before the deformation take place.

## IV. CONCLUSIONS

We studied the properties of ex-situ $MgB_2$ tapes and wires finding the optimal parameters to improve the $J_c$ for the high energy ball-milled tapes and for the wires realized at various synthesis temperature. We obtained the maximum $J_c$ value of $6 \times 10^4$ A/cm$^2$ corresponding to the optimal tape prepared with powders synthesized at 900°C and milled for 144h at 180rpm.

Similar results can be obtained for the tape milled for 72h at 240rpm then strongly reducing the milling time.

The best synthesis temperature is found to be 745°C, where $J_c$ presents a maximum. We speculate that this maximum is due to a balanced effect of unreacted magnesium content and the amount of reacted $MgB_2$ before the final sintering process at 920°C.